\documentstyle[pra,aps,epsf]{revtex}

\newcommand{\nn}{\noindent}
\newcommand{\beq}{\begin{equation}}
\newcommand{\eeq}{\end{equation}}
\newcommand{\bes}{\begin{eqnarray}}
\newcommand{\ees}{\end{eqnarray}}

\begin{document}
\draft
\thispagestyle{empty}
\title{
Constraints on non-Newtonian gravity 
from the Casimir force measurements between two crossed cylinders 
}
\author{
V.~M.~Mostepanenko\footnote{On leave from A.Friedmann Laboratory
for Theoretical  Physics, St.Petersburg, Russia.\\
Electronic  address: mostep@fisica.ufpb.br} and 
M.~Novello\footnote{Electronic  address: novello@lafex.cbpf.br}
}

\address
{
Centro Brasileiro de Pesquisas F\'{\i}sicas, Rua Dr.~Xavier Sigaud, 150\\
Urca 22290--180, Rio de Janeiro, RJ --- Brazil
}

\maketitle
\begin{abstract}
Constraints on the Yukawa-type corrections to Newtonian gravitational law
are obtained resulting from the measurement of the Casimir force between
two crossed cylinders. The new constraints are stronger than those 
previously derived in the interaction range between 1.5\,nm and 11\,nm.
The maximal strengthening in 300 times is achieved at 4.26\,nm.
Possible applications of the obtained results to the elementary
particle physics are discussed.
\end{abstract}


{
Non-Newtonian gravity is the subject of a considerable 
amount of literature
(see the monograph \cite{1} and references therein). Corrections to
Newtonian gravitational law at small distances are predicted by the
unified gauge theories, supersymmetry, supergravity and string theory.
They arise due to exchange of massless and light elementary particles
(such as axion, scalar axion, arion, dilaton, graviphoton etc) between
the atoms of two macrobodies \cite{2}. Another reason for non-Newtonian
gravity is the conceivable existence of extra spatial dimensions.
Recent trends show an increase in use of Kaluza-Klein field theories 
with a weak compactification scale \cite{3,4}. These theories 
predict that deviations from the Newtonian gravitational
law exist at submillimeter range \cite{5,6,7}. Namely in this range 
gravitational law is not confirmed experimentally (the common faith in
its validity up to the Planckean length of about $10^{-33}\,$cm is
nothing more than a far-ranging extrapolation \cite{8}).

Non-Newtonian gravity is usually described by the Yukawa- or power-type
corrections to the Newtonian potential which is inversely related to
distance. The constraints on the constants characterizing the magnitude
and interaction range of these corrections are usually obtained from
the Cavendish- and E\"{o}tvos-type experiments, Casimir and van der Waals
force measurements. The gravitational experiments are the best in the
interaction range
$10^{-2}\,\mbox{m}<\lambda<10^6\,$km
(see the most strong constraints obtained by this way in \cite{9}).
At submillimeter range the best constraints on the Yukawa-type
corrections to Newtonian gravity follow from the measurements of the
Casimir and  van der Waals forces \cite{2,10,10a}. Recently the new
experiments on measuring the Casimir force were performed
\cite{11,12,13,14}. With the results of \cite{11} the constraints on
the Yukawa-type interaction were strengthened in \cite{15} up to
30 times in the interaction range
$2.2\times 10^{-7}\,\mbox{m}\leq\lambda\leq 1.6\times 10^{-4}\,$m
(see also \cite{16}). By the use of \cite{12,13,14} the  
constraints known to date were strengthened in \cite{17,18,19}
 up to 4500 times in the interaction range
$4.3\times 10^{-9}\,\mbox{m}\leq\lambda\leq 1.5\times 10^{-7}\,$m. 
There are also other experiments (see, e.g., \cite{20}) but the constraints 
obtained from them are weaker than those mentioned above.

In paper \cite{21} the Casimir force was measured between the gold surfaces 
of two crossed cylinders covered by the thin layers of hydrocarbon.
Using a template-stripping method, the root mean square roughness of the
interacting surfaces was decreased up to 0.4\,nm. This gave the possibility
to measure the Casimir force at the closest separation of 20\,nm with
a resolution $\sim 10^{-8}\,$N. The decreased distance of closest
separation provides a way for obtaining more strong constraints on the
constants of Yukawa-type corrections to Newtonian gravity at small
distances. As shown in the present paper, 
the constraints, following from \cite{21},
are up to 300 times stronger than the previously known ones
in the interaction range
$1.5\times 10^{-9}\,\mbox{m}\leq\lambda\leq 11\times 10^{-9}\,$m.

In order to calculate non-Newtonian force between two crossed cylinders
we consider first two point masses $M_1$ and $M_2$ at a distance
$r_{12}$ apart. The effective potential of gravitational interaction
with account of Yukawa-type corrections is \cite{1,8,17,18}
\beq
V(r_{12})=-\frac{GM_1M_2}{r_{12}}\left(
1+\alpha_Ge^{-r_{12}/\lambda}\right).
\label{1}
\eeq
\nn
Here $G$ is Newtonian gravitational constant, $\alpha_G$ is
a dimensionless interaction constant, $\lambda$ is the interaction range.
The gravitational potential with account of power-type corrections is
of the form
\beq
V_n(r_{12})=-\frac{GM_1M_2}{r_{12}}
\left[1+\lambda_n^{G}\left(\frac{r_0}{r_{12}}\right)^{n-1}\right],
\label{1a}
\eeq
\noindent
where $r_0=10^{-15}\,$m is introduced to provide the proper 
dimensionality with different $n$ \cite{22a}, $\lambda_n^{G}$ is
the interaction constant depending on $n$. In both Eqs.~(\ref{1}),
(\ref{1a}) the non-Newtonian gravity dominates at small separations.
With increase of the separation the non-Newtonian corrections
vanish comparing the Newtonian contribution. 

The interaction potential acting between two crossed cylinders is obtained
by the integration of Eqs.~(\ref{1}), (\ref{1a}) over their volumes. 
The end effect consists of two contributions --- Newtonian one and
Yukawa- or power-type one. The case of
Yukawa-type interaction will be our initial concern. Newtonian gravitational
interaction will be estimated next. Let the first cylinder of radius
$R_1$ be arranged vertically so that its axis coincides with the axis $z$.
Both cylinders can be considered as infinite because their lengths are
much larger than their radii.

We consider an arbitrary point $P$ belonging to the second cylinder and 
from $P$ drop a perpendicular to the axis of the first cylinder. 
In a spherical coordinate system with center $O_1$ located at the 
intersection of the first cylinder axis and the above perpendicular the
coordinates of point $P$ are 
($r_2,\>\theta_2=\frac{\pi}{2},\>\varphi_2=\frac{\pi}{2}$). Here it is
suggested that $O_1P$ coincides with $y$ axis and angle $\varphi$ is
counted from $x$ axis. Distance $r_{12}$ is now a separation between
an arbitrary point of the first cylinder with coordinates
($r_1,\,\theta_1,\,\varphi_1$) and a point $P$.

Interaction potential of Yukawa-type between the first cylinder and a point
$P$ can be obtained by the integration over the cylinder volume
\beq
\Phi(r_2)=-GM_1M_2\alpha_GN_1
\int\limits_{0}^{2\pi}d\varphi_1
\int\limits_{0}^{\pi}\sin\theta_1\,d\theta_1
\int\limits_{0}^{R_1/\sin\theta_1}r_1^2\,dr_1
\frac{e^{-r_{12}/\lambda}}{r_{12}},
\label{2}
\eeq
\nn
where $N_1$ is the atomic density of the first cylinder material.

To calculate integrals in Eq.~(\ref{2}) let us take advantage of the 
expansion in spherical harmonics \cite{22}
\beq
\frac{1}{r_{12}}e^{-r_{12}/\lambda}=4\pi
\sum\limits_{l=0}^{\infty}a_l(r_1,r_2)
\sum\limits_{m=-l}^{l}Y_{lm}^{\star}(\theta_1,\varphi_1)
Y_{lm}(\theta_2,\varphi_2),
\label{3}
\eeq
\nn
where the coefficients are expressed in terms of Bessel functions
of imaginary argument
\beq
a_l(r_1,r_2)=\frac{1}{\sqrt{r_1r_2}}
I_{l+\frac{1}{2}}\left(\frac{r_{<}}{\lambda}\right)
K_{l+\frac{1}{2}}\left(\frac{r_{>}}{\lambda}\right),
\label{4}
\eeq
\nn
and $r_{<}\equiv\min(r_1,r_2)$, $r_{>}\equiv\max(r_1,r_2)$.

Now we take into account that the radii of the cylinders in \cite{21}
are rather large ($R_1=R_2=R=1\,$cm) comparing the values of $\lambda$
for which the strong constraints are obtainable at separations of
about 20\,nm. It follows from here, that not only $\lambda\ll r_{>}$
but also $\lambda\ll r_{<}$ (since only a thin layer adjacent to the
first cylinder surface contributes to the force). As a result the
conditions $r_{<}/\lambda,\,r_{>}/\lambda\gg 1$ are valid and one may use
the asymptotic expressions for Bessel functions of large arguments
\beq
I_{l+\frac{1}{2}}(z)\approx\frac{1}{\sqrt{2\pi z}}e^z,\quad
K_{l+\frac{1}{2}}(z)\approx\sqrt{\frac{\pi}{2z}}e^{-z}.
\label{5}
\eeq

Substitution of (\ref{5}) into (\ref{4}) in the case of $r_1<r_2$
leads to the following expression
\beq
a_l(r_1,r_2)=\frac{\lambda}{2r_1r_2}e^{(r_1-r_2)/\lambda}.
\label{6}
\eeq
\nn
For $r_1>r_2$ one obtains
\beq
a_l(r_1,r_2)=\frac{\lambda}{2r_1r_2}e^{(r_2-r_1)/\lambda}.
\label{7}
\eeq

Note that in both cases the value of $a_l$ under the above conditions 
does not depend on $l$. This gives the possibility to calculate explicitly
all integrals in Eq.~(\ref{2}). To do this we substitute (\ref{3}),
(\ref{6}), (\ref{7}) into (\ref{2}) and use the completeness relation
for spherical harmonics \cite{22}
\beq
\sum\limits_{l=0}^{\infty}
\sum\limits_{m=-l}^{l}Y_{lm}^{\star}(\theta_1,\varphi_1)
Y_{lm}(\theta_2,\varphi_2)=\delta(\varphi_1-\varphi_2)
\,\delta(\cos\theta_1-\cos\theta_2)
\label{8}
\eeq
\nn
at a point $\theta_2=\varphi_2=\frac{\pi}{2}$. The result is
\beq
\Phi(r_2)=-2\pi G\rho_1M_2\alpha_GR_1\lambda^2e^{R_1/\lambda}
\frac{1}{r_2}e^{-r_2/\lambda},
\label{9}
\eeq
\nn
where $\rho_1=M_1N_1$ is the density of the first cylinder.

In order to calculate the interaction potential between the two cylinders
it is necessary now to integrate Eq.~(\ref{9}) over the volume of the
second cylinder which is aligned perpendicular to the first one at the
closest separation $a$. Thus, the separation between the axes of the
cylinders is $R_1+R_2+a$. 
We consider now the cylindrical coordinate system. Let $z$ axis of it
coincides with the axis of the first cylinder and let the origin be at
a point nearest to the axis of the second cylinder. In this coordinate
system the coordinates of any point belonging to the second cylinder
are
(${\tilde r}_2,\,{\tilde\varphi}_2,\,{\tilde z}_2$).
Here ${\tilde r}_2$ is the separation between the point of the second
cylinder and the axis of the first one.

Interaction potential between the two cylinders is given by
\beq
U(a)=4N_2
\int\limits_{0}^{\pi/2}d{\tilde\varphi}_2
\int\limits_{0}^{R_2}d{\tilde z}_2
\int\limits_{\rho_{\min}({\tilde z}_2,
{\tilde\varphi}_2)}^{\rho_{\max}({\tilde z}_2,
{\tilde\varphi}_2)}
d{\tilde r}_2\,{\tilde r}_2
\Phi({\tilde r}_{2}).
\label{10}
\eeq
\nn
Here the function $\Phi$ is defined in (\ref{9}) and 
\beq
\rho_{\max(\min)}({\tilde z}_2,{\tilde\varphi}_2)=
\frac{R_1+a+R_2\pm\sqrt{R_2^2-{\tilde z}_2^2}}{\cos{\tilde\varphi}_2}.
\label{10a}
\eeq

Calculating integrals in (\ref{10}) with the help of \cite{23a} and
using the conditions $\lambda,\,a\ll R_1,\,R_2$ one obtains
\beq
U(a)=-4\pi^2G\rho_1\rho_2\alpha_G\lambda^4\sqrt{R_1R_2}
e^{-a/\lambda},
\label{11}
\eeq
\nn
where $\rho_2$ is the density of the second cylinder.

The Yukawa-type force acting between the two cylinders is obtained
from (\ref{11})
\beq
F_Y(a)=-\frac{dU(a)}{da}= -4\pi^2G\rho^2\alpha_G
\lambda^3\sqrt{R_1R_2}e^{-a/\lambda}.
\label{12}
\eeq
\nn
Exactly the same expression can be obtained from the Yukawa-type
energy of two plane parallel plates by the application of proximity
force theorem \cite{23b}. As for the power-type interactions,
distinct results are obtained with different $n$. By way of
example, with $n\geq 5$ the power-type force acting between
two crossed cylinders can be represented as
\beq
F_n(a)=-\frac{4\pi^2r_0^{n-1}G\lambda_n^{G}\rho^2}{(n-2)(n-3)(n-4)}
\frac{R}{a^{n-4}}.
\label{13a}
\eeq

In the experiment \cite{21} two identical cylinders where made of
the same material and covered by the thin layers of gold (of thickness
$\Delta_1$ and density $\rho^{\prime}$) and hydrocarbon (thickness
$\Delta_2$, density $\rho^{\prime\prime}$). The force acting in 
experimental configuration can be easily obtained by the combination
of several expressions of Eqs.~(\ref{12}), (\ref{13a}). 
For the Yukawa-type force the result is
($\rho=\rho_1=\rho_2$)
\beq
F_Y(a)=-4\pi^2G\alpha_G\lambda^3Re^{-a/\lambda}
\left[\rho^{\prime\prime}\left(1-e^{-\Delta_2/\lambda}\right)
+\rho^{\prime}\left(1-e^{-\Delta_1/\lambda}\right)e^{-\Delta_2/\lambda}+
\rho e^{-(\Delta_1+\Delta_2)/\lambda}\right]^2.
\label{13}
\eeq

Returning back to the Newtonian gravitational force, given by the first 
term in the right-hand side of Eqs.~(\ref{1}), (\ref{1a}),
we find it negligible 
to compare with the Casimir force at the closest separations. Actually, 
the gravitational force acting between the crossed cylinders spaced at 
$a\ll R$ apart can be roughly estimated as the force acting between
two spheres of radius $R$
\beq
F_N\sim G\frac{\left(\frac{4}{3}\pi\rho R^3\right)^2}{4R^2}=
\frac{4}{9}\pi^2G\rho^2R^4.
\label{14}
\eeq
\nn
If the cylinders are made of quartz with 
$\rho=2.23\times 10^3\,$kg/m${}^3$ the Newtonian gravitational force is
estimated as $F_N\sim 1.4\times 10^{-11}\,$N. The Newtonian force is
rather small, even though the cylinders as a whole are made of gold
with  $\rho^{\prime}=18.88\times 10^3\,$kg/m${}^3$.
In this case $F_N\sim 1\times 10^{-9}\,$N which is still much less
than the Casimir force at the separations of 20\,nm equal to
$F_C\approx 9.4\,\mu\mbox{N}/\mbox{m}\times 2\pi R\approx 6\times 10^{-7}\,$N
\cite{21}. As a consequence one should not add the contribution of 
Newtonian gravity to the non-Newtonian terms given by Eqs.~(\ref{12}),
(\ref{13a}). Newtonian gravity can be neglected  
when obtaining constraints on hypothetical interactions 
from the results of Casimir force measurements at
smallest separations.

We are coming now to the obtaining
constraints on the Yukawa-type interaction.
According to the result of \cite{21} the theoretical expression for
the Casimir force acting between two crossed cylinders was confirmed 
within the limits of experimental error of force measurement 
$\Delta F=10\,$nN. This theoretical expression was obtained with regard 
to the corrections due to the finite conductivity of the boundary metal,
stochastic roughness, covering hydrocarbon layer, and nonzero temperature
(in fact temperature corrections are insignificant at the separations
$a<100\,$nm considered in \cite{21}). The relative accuracy of force
measurements at the smallest separations can be estimated as
$\delta=\Delta F/F_C\approx 1.7$\%. It increases quickly, however, with
increasing of $a$.

Since no any Yukawa-type interaction was observed in the limits of 
experimental error, it is evident that
\beq
\left\vert F_Y(a)\right\vert\leq\Delta F,
\label{15}
\eeq
\nn
where $F_Y(a)$ is defined by Eq.~(\ref{13}).

The constraints on the parameters of Yukawa-type interaction
$\alpha_G$ and $\lambda$ are obtained from Eq.~(\ref{15}) by
substituting the values of quartz and gold densities given above
and also the density of hydrocarbon layer
$\rho^{\prime\prime}\approx 0.85\times 10^3\,$kg/m${}^3$.
The thicknesses of the layers are $\Delta_1=200\,$nm, $\Delta_2=2.1\,$nm.
The strongest constraints follow at the closest separation distance 
$a=20\,$nm. The computational results are shown in Fig.~1 by the curve 1 
at the logarithmic scale. In the same figure the curves 2 and 3
show the constraints which follow \cite{18,19} from the Casimir
force measurements between a plane disk and a spherical lens by means
of atomic force microscope \cite{13,14}. Curve 4 presents the constraints
obtained from the Casimir force measurements between dielectrics \cite{2}.
Curve 5 shows constraints following from the measurements of the van der
Waals force \cite{10}. The regions above the curves 1--5 are prohibited
by the results of the corresponding experiment. The regions below the 
curves are permitted.

As seen from Fig.~1, the Casimir force measurements of \cite{21} lead
to strongest constraints on the constants 
of Yukawa-type corrections to Newtonian
gravitational law within the interaction range
$1.5\,\mbox{nm}<\lambda <11\,$nm. The largest strengthening in 300 times
is achieved at $\lambda=4.26\,$nm. For $\lambda <1.5\,$nm the best
constraints follow from the measurement of the van der Waals force.
For $11\,\mbox{nm}<\lambda <150\,$nm the constraints 
from the Casimir force measurements
between gold surfaces of a spherical lens and a disk \cite{14} are 
the strongest ones.

According to Fig.~1, at nanometer range the Yukawa-type corrections are 
still permitted by the experiment which are in excess of the Newtonian
gravitational interaction of more than 20 orders of magnitude.
The nanometer range corrections to Newtonian gravity are of especial
interest for the weak-scale compactification schemes with three extra
spatial dimensions (total space-time dimensionality $N=7$). In fact,
in this case the compactification dimension is equal to \cite{23}
\beq
R\sim10^{\frac{30}{N-4}-19}\,\mbox{m}=10^{-9}\,\mbox{m}=
1\,\mbox{nm},
\label{16}
\eeq
\nn
and at the several times larger separations the non-Newtonian gravity 
should be noticeable. Exchange of the hypothetical particles
mentioned above also may contribute to the Yukawa-type
force. That is the reason why the further strengthening of constraints
on the corrections to Newtonian gravity in nanometer range 
presents interest for the elementary particle physics.

It is easy to check that no strong constraints are obtainable from
the experiment \cite{21} on the constants of power-type corrections
to Newtonian gravity. For $n=5$, as an example, the constraint
following from Eq.~(\ref{13a}) and an inequality
\beq
|F_5(a)|\leq\Delta F=10\,\mbox{nN}
\label{17a}
\eeq
\noindent
is $|\lambda_5^{G}|\leq 0.92\times 10^{49}$. This is much weaker
than $|\lambda_5^{G}|\leq 2.1\times 10^{47}$ obtained from the
Cavendish-type experiments \cite{27}. As noted in \cite{15},
the Casimir force measurements between metallic surfaces produce
weaker constraints on the power-type interactions
than the gravitational experiments (see \cite{9} for the latest
results).

As was shown above, the Casimir force measurement between the metallized
surfaces of two crossed cylinders gives the possibility to strengthen
constraints on the non-Newtonian gravity 
of Yukawa-type up to 300 times in the
nanometer range. During the last few years several new measurements
of the Casimir force between metallic bodies were performed
\cite{11,12,13,14,21} and each of them gave possibility to obtain more
strong constraints on the constants of Yukawa-type corrections to
the Newtonian gravitational law \cite{15,16,17,18,19}. Notice that all
these experiments were designed for the registration of the Casimir
force only. There were no opportunities used 
especially in order to increase
the sensitivity of an experiment to the probable Yukawa-type interaction
which were proposed in the literature (see, e.g., \cite{2,8}). This 
means that the potentialities of the Casimir force measurements have not
been exhausted.

Currently several new experiments on the measurement of the
Casimir force are in preparation. They will give possibility to
strengthen further the constraints on non-Newtonian gravity up to $10^4$
times in a wide interaction range from $10^{-9}\,$m till $10^{-3}\,$m.
As a result the values of $\alpha_G<0.1$ will be achieved near the right
boundary of this interval (the method using the dynamical Casimir force
\cite{20} is also high promising in the process). In this way the
Casimir effect is quite competitive with the modern accelerators and
also with the E\"{o}tvos- and Cavendish-type experiments in the search
for light elementary particles and hypothetical long-range
interactions predicted by the modern theories of fundamental
interactions. 

The authors are grateful to T.~Ederth, G.~L.~Klimchitskaya 
and D.~E.~Krause
for helpful discussions. 
They acknowledge the partial financial
support from FAPERJ and CNPq.

}
\newpage
\widetext
\begin{figure}[h]
\epsfxsize=15cm\centerline{\epsffile{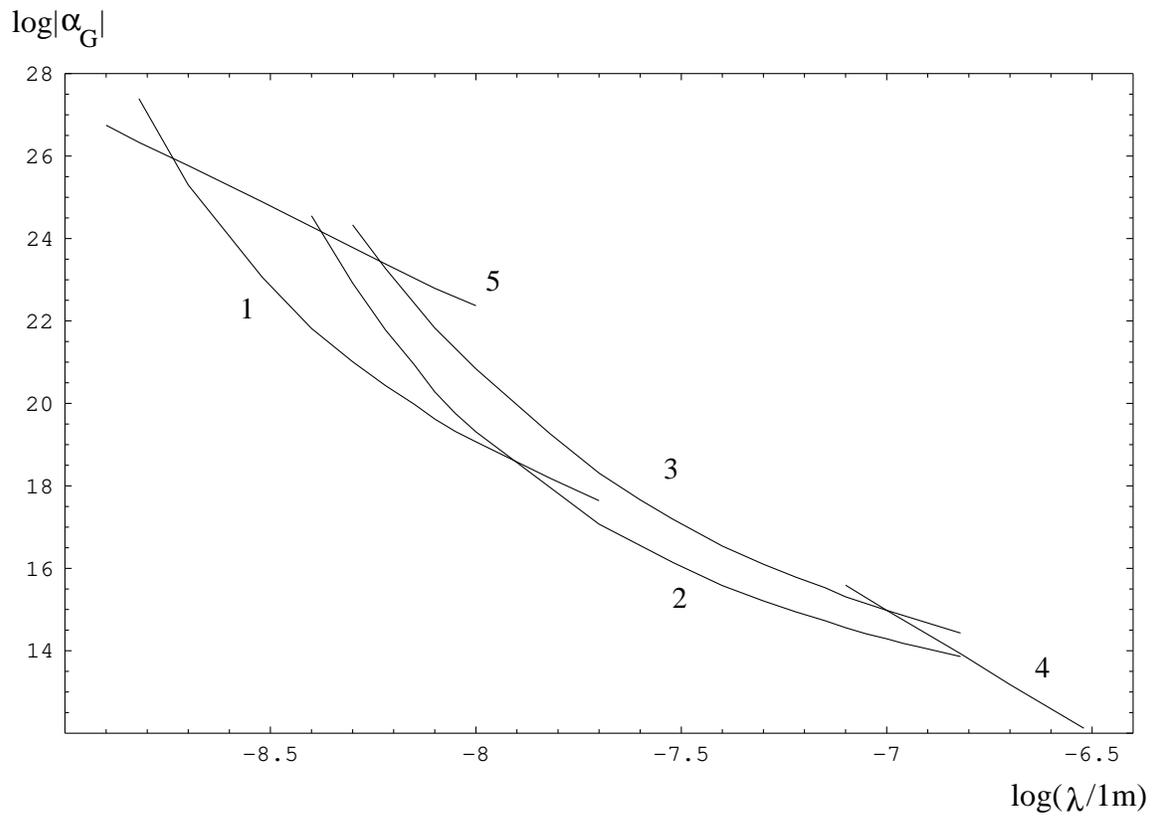} }
\vspace*{1cm}
\caption{
Constraints on the constants of Yukawa-type  corrections
to Newtonian gravity.
Curves 4 and 5 were obtained
from the measurements of the Casimir and van der Waals forces   
respectively between dielectrics.
Curves 2 and 3  follow from the  Casimir force measurements  
between gold and aluminum surfaces by means of
 atomic force microscope.
Curve 1 is obtained in this paper from the Casimir force
measurement between two crossed cylinders.
The regions below the curves are permitted, and those above the 
curves are prohibited.
}
\end{figure}
\end{document}